# Theatrical Language Processing: Exploring AI-Augmented Improvisational Acting and Scriptwriting with LLMs

**Sora Kang, Joonhwan Lee**
HCI+D Lab, Seoul National University
Seoul, Korea
sorakang@snu.ac.kr, joonhwan@snu.ac.kr

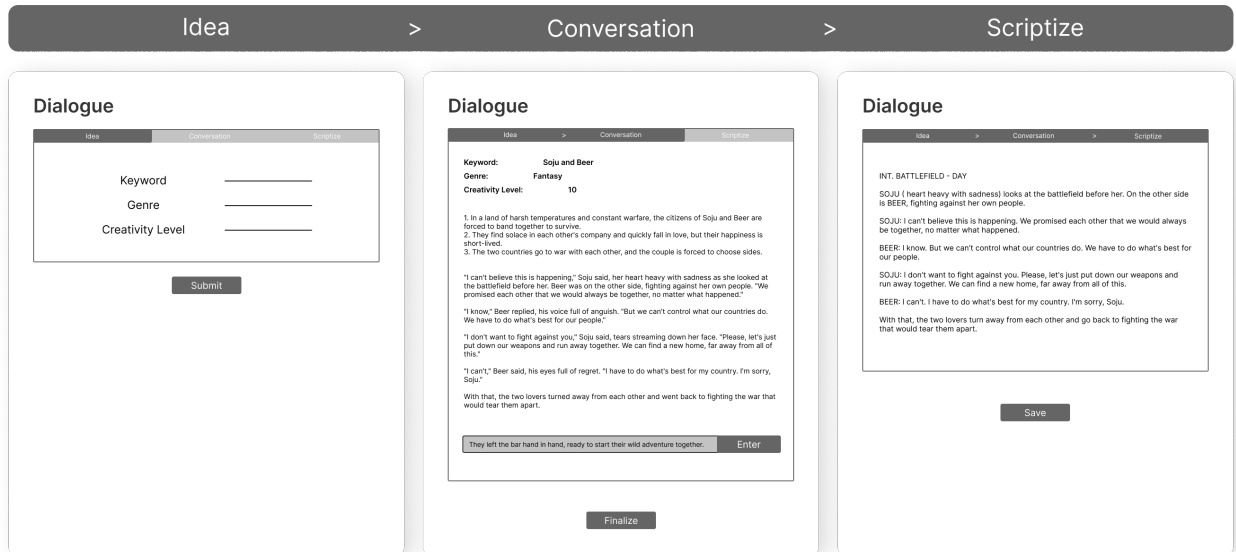

Figure 1: Final UI system design and graphic component for Dialogue function

## Abstract

The increasing convergence of artificial intelligence has opened new avenues, including its emerging role in enhancing creativity. It is reshaping traditional creative practices such as actor improvisation, which often struggles with predictable patterns, limited interaction, and a lack of engaging stimuli. In this paper, we introduce a new concept, Theatrical Language Processing (TLP), and an AI-driven creativity support tool, Scribble.ai, designed to augment actors' creative expression and spontaneity through interactive practice. We conducted a user study involving tests and interviews with fourteen participants. Our findings indicate that: (1) Actors expanded their creativity when faced with AI-produced irregular scenarios; (2) The AI's unpredictability heightened their problem-solving skills, specifically in interpreting unfamiliar situations; (3) However, AI often generated excessively detailed scripts, which limited interpretive freedom and hindered subtext exploration. Based on these findings, we discuss the new potential in enhancing creative expressions in film and theater studies through an AI-driven tool.

## Introduction

We live in an era where technology and artificial intelligence have become cornerstones of various fields. Particularly, Natural Language Processing (NLP)-based tools have been increasingly utilized, transforming the landscape of numerous industries with their potential to understand, interpret, and recreate human language in an intelligible and meaningful manner. The trajectory of AI tool development has shifted from purely productivity enhancement to encouraging creative processes. Rather than automating mundane tasks to augment organizational productivity, the spotlight is now on augmenting human creativity – a distinguishing human trait that allows us to imagine, invent, and innovate. The recent creative support tools display how AI can help humans think outside the box, ignite untapped creativity, and push the boundaries of innovation.

Actors, in particular, represent a demographic for whom creativity is an essential currency. There have been many different definitions of 'acting', but they all converge on one core idea: "Acting is doing not pretending—real doing" [1,3]. The nature of play scripts demands that actors create the additional details that make characters believable. In particular, since the script does not provide all the information needed, one of the essential roles of the actor is to create the context and subtext [12:7]. Hence, 'creativity' is a key skill that actors require in order to involve entering a creative state of mind, allowing actors to "live truthfully under imaginary circumstances" [9].

In the interest of fostering enhanced creativity and spontaneity among actors, our focus is on improvisational acting. Improv is an unscripted or not fully unscripted form of theater where performers spontaneously create the storyline and characters. Known for being a rewarding and effective mechanism to boost creativity, it allows the actors to think on their feet and create in the moment, fostering a sense of spontaneity that leads to more genuine and vibrant performances. This helps actors live and breathe their characters on stage, preventing an all-too-common problem of "mannerism" where actors become too familiar with their scripted situations. This is precisely where AI assumes a pivotal role. By generating an indefinite number of situational scenarios and scripts catering to numerous contextual nuances, it makes improv practice more accessible and dynamic. In this context, we introduce a new concept named 'Theatrical Language Processing', a derivative of NLP specifically catered to the needs of the theater.

To demonstrate the applicability of TLP, we developed 'Scribble.ai', an AI-driven creativity support tool for actors. The tool aims to challenge the actors' creative boundaries, encourage novel interpretations, and assist in honing their performative skills. We aspire to bolster their inherent creativity, enrich their craft, and ultimately, enhance their performances. It is envisaged that this tool could serve as an interactive partner during their practice sessions, infusing them with the unpredictability characteristic of human interactions. This paper focuses on elucidating the design and functionality of this tool and validates its efficacy in augmenting creativity among actors. We conducted a comprehensive user study involving pre-interviews, interactive exercises, and post-activity interviews to evaluate its impact on enhancing creativity. Our sample included 14 participants: 10 theatrical actors and 4 directors/acting coaches. Participants were tasked with understanding random scripts, and subsequent interviews helped us identify the unique characteristics of human and AI authorship. In the second phase, Scribble.ai was used to create real-time scripts and interact with improv acting. Following these activities, we conducted post-activity interviews.

The results of the study indicated the following:
1. Participants demonstrate heightened creativity when engaging with AI-produced scenarios.
2. A certain degree of unpredictability imparted by AI content positively impacts actors' problem-solving abilities, particularly with regards to understanding and engaging with unfamiliar contexts and language.
3. However, AI's tendency to produce overly detailed content could limit the participants' creative freedom and impede opportunities for interpretation and sub-text exploration.

This research highlights the significance of integrating theater and HCI studies, both of which examine human interaction—one in the physical realm, the other digitally. Their convergence offers innovative approaches to improvisational theater and the creation of interactive, AI-driven platforms for actors. We believe our study emphasizes the growing necessity for interdisciplinary research in these areas. This cross-disciplinary research is rooted in the intersection of HCI and theater studies. As both HCI and theater involve human interaction and expression [6,7], the research becomes notably impactful within the field of HCI. We contend that interdisciplinary research serves to strengthen our understanding of humans' complex interactions and behaviors, thereby underscoring the critical need for more extensive studies at the crossroads of HCI and theater studies.

## Related Work

### Improvisational theater

Improvisational theater(improv) is the type of theater where the majority or all of what is performed is spontaneously created by its actors on stages without a script or plan [15,21,30]. This has been often used in acting training for stage, film, and television to help actors develop their skills in reacting to the unexpected, creating characters spontaneously, and working together as an ensemble. The earliest well-recognized use of improvisational theater can be found in the Atellan Farce–a precursor to Roman comedy–in 391 BC [24]. In the 1890s, Stanislavski–a founder of acting theory–devoted a great deal of attention to improvisation in training and rehearsal in his books such as 'An Actor Prepares'[20]. Between the 1940s and 1960s, Viola Spolin made another significant contribution by further developing improv activities, which she codified in her book 'Improvisation for The Theater', which is the first book to give specific methods for learning to do and teach improv [19].

One of the most popular improvisation methods used to train students or acting trainees as well as professional actors developed by Konstantin Stanislavsky is étude, originally meaning 'study' in French. As a theatrical term, it is used to mean a short script or description that only provides the beginning part of a situation without an ending so that the actors can fill in the story using their instincts and imagination and drive it to the end [25]. More recently, an online form of improv theater known as 'Netprov' has been developed. Netprov, short for networked improv, makes scenarios that are networked, collaborative, and improvised on a real-time basis [26]. These studies show the great importance of improvisation for actors by examining how improv was invented and has evolved throughout history.

### Creativity support tools

Creativity is normally defined as the ability to generate new concepts in line with the context where they are realized [2]. Over the past 40 years, computing experts have successfully built productivity support tools for users to accomplish their tasks more quickly, effectively, and with fewer mistakes. Now, they are focusing on developing creativity support tools (CST) that allow people to interact, explore, ideate, and envision [18]. The goal of CST is to get more people to be more creative more often so they can effectively deal with a wider range of obstacles across various fields [17].

For CST in writing, 'Servicescape writing prompt', a writing prompt generator, is designed to be used for writing inspiration, and it can provide writing ideas across many genres, from Fantasy to Science Fiction to Horror to Romance [27]. Users select their desired themes, and the tool provides a short prompt to inspire their writing. However, it is limited by a finite set of built-in prompts, leading to repetitive suggestions over time. Another tool, 'Jasper', serves as an AI copywriter. Utilizing artificial intelligence and machine learning, Jasper creates content for blog posts, landing pages, social media, ads, and marketing emails [28]. It also has merit in that it is able to generate original content for marketing purposes. The third example of a writing tool is an AI storytelling tool, 'Sudowrite'. Sudowrite is primarily designed for novel writers by Amit Gupta. Given a passage of text by a user, Sudowrite generates suggestions for what to write next [5]. A large number of existing tools have shown the effectiveness and potential of AI usage in writing. However, the existing tools are almost exclusively focused on marketing purposes or writing skill improvement, and it seems likely that they really support productivity rather than creativity.

## Natural Language Processing (NLP) and Theatrical Language Processing (TLP)

In this paper, we propose the new terminology 'Theatrical Language Processing (TLP)', inspired by Natural Language Processing (NLP). NLP is defined as a theoretical framework of computational techniques for analyzing and representing naturally occurring texts in order to generate human-like language [8]. The related studies pertinent to our explorative journey into AI-driven theatrical language processing can be broadly categorized into two primary areas: NLP for script and theater play generation; and Human-AI collaboration in creating scripts and enhancing performance.

NLP's application in theater has been the subject of various studies, each with distinct methodologies and results. Efforts have been made in automatic theater play script generation using deep generative language models and hierarchical generation methods [13]. Researchers have also developed models for movie scripts by analyzing textual data from diverse movie elements [29]. Progress in computer-assisted scriptwriting demonstrates the ability to extract insights from natural language narratives, aligning with storylines or characters [14]. Moreover, integrating NLP with other digital technologies in theater enhances the semiotics of presentations and enriches audience experiences [22].

In terms of human-AI collaboration, recent research has begun to explore the potential of leveraging AI as a co-author. Studies have explored AI solutions to writer's block, showing promise in inspiring creative writing [31,32]. The Wordcraft model further illustrates this potential by employing Large Language Models (LLMs) to develop an editor interface that promotes continuous writing, requests additional details, or suggests text revisions [23]. These studies indicated that even though AI lacks long-range semantic coherence, its usefulness for long-form creative writing can be greatly enhanced by hierarchical application [10,11]. By using prompt chaining to build a structural context, AI systems could generate coherent scripts complete with required elements such as title, characters, story beats, location descriptions, and dialogue [11].

Despite significant contributions by these studies, most studies have solely focused on script generation designed to aid scriptwriters, and there is little research considering the perspective of actors. Furthermore, while some studies have begun to explore the co-creative processes between AI and humans, the focus has been largely on maintaining long-range semantic coherence, rather than exploration of creativity possibilities. Addressing these limitations, our research proposes 'Theatrical Language Processing (TLP), the new transdisciplinary field of linguistics, theatrology, computer science, and artificial intelligence which deals with the interactions between computers and theatrical language.

In this paper, we seek to narrow down our focus to the intersection of improvisational theater, creativity support tools, and theatrical language processing and examine how to combine improv practices with AI technology to encourage and facilitate creativity. Therefore, we will highlight and analyze prior studies that specifically deal with improvisational theater and creativity support tools and discuss the potential of incorporating TLP into this context.

## Scribble.ai

### System Description

'Scribble.ai' is an AI-driven creativity support tool for actors where users can input their ideas for the storyline, and the AI generates the script based on user input. To be specific, this tool provides the improv-purposed scripting for 1) a real-time interactive dialogue and 2) a monologue, and it requires either mandatory or optional inputs.

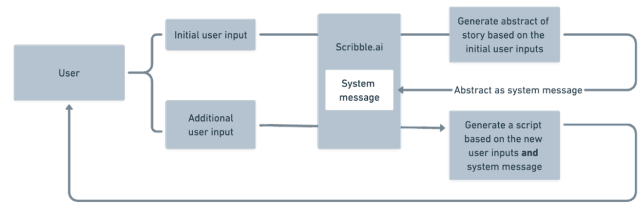

Figure 2: Interaction system of Scribble.ai

The interaction system (see Figure 2) requires clear objectives for both the user and AI within the tool. The user's goal is to quickly and conveniently acquire a well-crafted script for improvisational practice, while the AI's task is to comprehend the user's language, process it, and generate a pertinent script based on their input. However, a typical challenge encountered in GPT-3.5 projects is that the initial sentences are well-crafted, but subsequent ones often fall short in maintaining coherence, as noted by Ben Shneiderman's feedback [33]. Further, when crafting stories, maintaining the central topic is vital; regardless of the thrilling and dynamic events occurring within the narrative, the main topic must remain intact. To ensure this, we have created an innovative approach where we initially formulate a brief abstract based on user input. This abstract is then transformed into system prompts, which ensures even if new user inputs are added or the content lengthens, the story is produced grounded in the centered topic.

**Function**

**Interactive Dialogue Generation**
The dialogue generation function consists of three stages: Idea Input, Conversation, Scriptizing. To break down each stage, there are three required inputs in the 'Idea Input' stage - keywords, genre, and randomness.

Based on the three inputs, AI will generate a script with a short summary and dialogue and show the output as a result on the interface. Each input refers to the following.
- Keyword (string type): Topic words that will act as a focus for AI to generate the storyline, e.g. orange, summer, iPhone, etc.
- Genre (string type): Categories that will determine the style, characters, setting, and tone of the story, e.g. Horror, Sci-fi, Romance, etc.
- Creativity Level (float type, 0-1): An elevated level of randomness infuses ambiguity and challenge into the generated script; more randomness means more chance of unfamiliar story progression.

In the next stage 'Conversation', users can input a new line or new prompt in case they want to further develop the story with AI. For example, users can simply type in the textbox "Sara: I want to go home", or "The girl named 'Jessica' comes and try to talk to the people", and the tool will generate the next part of the dialogue based on the new input. It allows users to interact with the tool on a real-time basis in the form of conversation. This feature is a result of the valuable input provided by Terry Winograd and Eric Paulos, keeping the narrative fluid, engaging and user-guided [34,36]. The system allows for variables such as sudden changes ("Introduce a dragon") to be swiftly integrated.

In the last stage 'Scriptzing', a user can finalize their scripts and reformat them into a screenplay format. Drawing on the conceptual approach put forward by Terry Winograd [36], the focus should not solely be placed on the algorithm construction; instead, one must understand the context and think about the people, not the technology. Therefore, it is crucial that our end-users, acting professionals in this case, find comfort in using 'Scribble.ai.' To this end, Scribble.ai has been constructed to generate scripts tailored to user intent and theatrical needs, ensuring that the theater-related professionals, who are our primary users, find the system convenient and accessible. Additionally, to optimize the system so that it harmoniously integrates with complex theatrical elements, the tool offers an export feature that compiles user inputs and generated texts, including the story summary and finalized scripts, into a .txt file for easy access.

**Monologue Generation**
This tool also provides a monologue writing function based on three user inputs - one sentence, emotion, and randomness level. Each input refers to the following, and Figure 3 shows the final UI system design and graphic component for the monologue function.
- One sentence (string type): The main sentence that AI uses as context for the monologue, e.g. Emily regrets that she dropped the school
- Emotion (string type): The words that will determine the speaker's circumstances, mood, or relationships with others. There can be multiple or mixed emotions, e.g. depressed, excited, sad but happy, etc.
- Creativity Level (same as above in dialogue function)

**User Interaction flow**

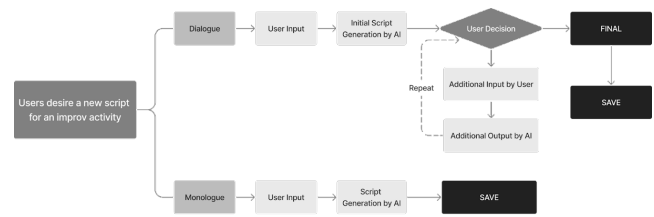

Figure 3: Final user interaction flow

**Dialogue**
1. A user desires a new dialogue script for an improv.
2. The user provides three contextual inputs for AI and presses the **Submit** button.
3. The AI generates the initial script based on the user input and prints it onto interface.
4. The user decides to continue the storyline or finalize it.
5. Opting to extend the scene, the user writes a new prompt in the textbox and presses the **Enter** button
6. The AI continues the storyline based on the new prompt and prints additional text below the current script.
7. Steps 4-6 repeat until the user is satisfied with the result
8. The user finalizes the script by pressing the **Finalize** button.
9. The AI reformats the script and prints the completed version below the unformatted script in a new section.
10. The user saves the finalized script file as a .txt by pressing the **Save** button.
11. The user uses the script for the improv activity.

**Monologue**
1. A user desires a new monologue script for individual improv training.
2. The user provides three contextual inputs for AI and presses the **Submit** button.
3. The AI generates a script based on the user input and prints it onto the interface.

4. The user saves the script file as a .txt by pressing the **Save** button.
5. The user uses the script for individual improv training.

The algorithm below shows the pseudocode of the interactive system of this tool.

ALGORITHM 1: Iterative Algorithm

```
IMPORT Anvil.Server as Frontend
IMPORT Google.Colab as Backend

CLASS Dialogue {
        Keyword, Genre, NewPrompt: user input, string type
        Randomness: user input, float type(0-1)
        Script, FinalScript: string type

        FUNCTION IdeaInput {
    Script = InitialScript(Keyword, Genre, Randomness) }
        FUNCTION Conversation {
    Script = Script + ConversationFollowup(NewPrompt) }
        FUNCTION Scriptizing {
    FinalScript = Reformat(Script) }
        FUNCTION Save {
    generate .txt file with FinalScript }
}

CLASS Monologue {
        OneSentence, Emotion: user input, string type
        Randomness: user input, float type(0-1)
        Script: string type

        FUNCTION Monologue {
    Script = MonoWriting(Keyword, Genre, Randomness) }
        FUNCTION Save {
    generate .txt file with Script }
}
```

## Study Design

In order to evaluate the efficacy of Scribble.ai as a Creativity Support Tool for actors, we designed a study consisting of two types of tests, follow-up interviews.

## Participants

We recruited 14 participants, including 10 actors and 4 directors and acting teachers from the same theater company. The group's mean age was 30.78 and the SD was 7.51, (Director/Teacher: Mean = 37.75, SD = 6.50, Actor: Mean = 28.38, SD = 4.13, Theater and Film Student: Mean= 21, SD = 1). As recommended by Ben Shneiderman, who originated the concept of Creativity Support Tools, we included directors and acting teachers in the study to obtain feedback on how the tools seem to enhance the actors' creativity since they are the experts appropriately equipped to evaluate the actors' performances [33]. We further reasoned that since directors are familiar with the usual performances of their actors, they could provide a comparative analysis of performance variations when AI is used. Being familiar with the standard acting patterns of the actors, they can discern even subtle deviations and enhancements from the usual performances. Prior to the study, we fully briefed all participants on the purpose and procedures, and we provided a step-by-step guide document to explain how Scribble.ai works. The broad portrait of our participants was as follows:

| Number | Ages | Gender | Occupation |
| --- | --- | --- | --- |
| P1 | 32 | Male | Actor/Director |
| P2 | 34 | Male | Director |
| P3 | 39 | Male | Director/Acting Teacher |
| P4 | 46 | Female | Director/Acting Teacher |
| P5 | 28 | Female | Professional Actor |
| P6 | 31 | Male | Professional Actor |
| P7 | 21 | Female | Theater and Film Student |
| P8 | 29 | Male | Professional Actor |
| P9 | 33 | Female | Professional Actor |
| P10 | 27 | Male | Professional Actor |
| P11 | 30 | Female | Professional Actor |
| P12 | 22 | Male | Theater and Film Student |
| P13 | 26 | Female | Professional Actor |
| P14 | 20 | Female | Theater and Film Student |

### Task 1. Human vs AI-authored scripts interview

Three human-authored scripts and three Scribble.ai-authored scripts were provided in random order and given 10 minutes to analyze them. We then interviewed them about how they distinguished between human and AI authorship to better understand participants' perception towards AI.

### Task 2. Post-activity interview after utilizing Scribble.ai in real-time

Participants utilized Scribble.ai in real-time to generate scripts, act out, and interact with given improvisations. Subsequent to the activity, an interview was conducted to gather qualitative feedback on their experiences and suggestions for improvement. The directors' reviews entailed their observations on the actors' adaptability, creativity, and problem-solving abilities during the interfaces with Scribble.ai.

### Findings

We delved into participants' perceptions, identifying trends and insights behind their responses. Notably, we found that they preferred distinct differences between human-written scripts and those created by AI. Participants appreciated the detail and clarity of AI scripts, yet they also appreciated the human scripts' flexibility and interpretative openness. We also explored real-time experiences of improvisation acting using Scribble.ai and participants' general considerations about interacting within an AI environment.

*Clarity vs Creativity - Heavy Detail or Room for Interpretation*: Participants appreciated the details and clarity in the AI scripts, which made scenario interpretation easier for actors. However, the counterpart human scripts allowed for more interpretational freedom, fostering creativity in the process. Human scripts were found to be abstract and open-ended, inviting diverse interpretation depending on the context. "The AI-written script was detailed to the point that it eliminated the need for guesswork, which admittedly made the acting process easier, but I'm not sure if that's necessarily a good thing." P9 said.

*Engagement Over Time: Depth of Story Development*: While AI scripts tended to entice with interesting themes and initial scenarios, they lacked depth and conflict as dialogue progressed. The depth of human-authored scripts may not have been evident at the outset but became more engaging due to their detailed settings and descriptive scenarios. As dialogue evolved, participants observed a growing tension that culminated in a captivating climax.

*Perception of Errors and Vocabulary: Participants' Stereotypical Views*: Participants erroneously tied typos and grammatical errors to human-written scripts, assuming the AI would not make such mistakes. Interestingly, when encountering difficult or archaic vocabulary, participants tended to categorize these as AI-generated. "There are typos in this script. It must be written by a human." P12 said.

*Improvisation Acting Activity: Participants' Reflections*: Participants' desire to make every decision was evident. They found the training effective for improving acting skills. Moreover, they reported that improvisation acting with AI was a fun and unique experience. They saw value in the unique themes and challenging situations presented by the AI, forcing them to expand their imagination and creativity.

*Areas for Improvement*: Despite the generally positive feedback, participants identified areas needing improvement. Increasing the AI script's logical flow, providing better user guidance, voice-interaction, and allowing other inputs than genre and keywords were areas of concern. These findings provide valuable feedback for further development and improvement of AI's role in artistic spaces, specifically in contributing to and enhancing human creativity in real time.

In the assessment of Scribble.ai as a creativity support tool, directors and acting coaches-being best positioned to evaluate actors' performances- indicated that the AI tool appeared to enhance the actors' creative abilities in several ways as follows.

*Enhanced imagination and creativity*: Initially, actors were puzzled by unconventional topics and sequences generated by the AI. However, as the training proceeded, there was a noticeable broadening in the actors' imaginative and creative capacities, evidenced by their ability to adeptly adjust to these unusual scenarios. Notably, they succeeded in creatively bridging connections within a seemingly disjointed scenario-lack of dialogue continuity-and managed to act it out by envisaging the implied situations between lines.

*Problem-solving skills and adaptability*: The unconventional topics and sequences produced by AI compelled actors to grapple with unanticipated problems. However, through these challenges, actors demonstrated competency in identifying unique problem-resolution strategies and adapting to novel situations. For instance, despite initial confusion with scripts set in complex and unfamiliar environments, actors reinterpreted these scenarios creatively, expanding their problem-solving skills and emotional range.

Conversely, the study also revealed potential limitations associated with the AI scripts. Scripts that provide excessively specific directives or straightforwardly stipulate emotional states, such as "I'm feeling sad now," occasionally hinder actors in their pursuit of exploring unique emotional interpretations and expressions. Actor P6 noted that overly directive scripts, though easier to follow, limited his ability to interpret the plot and create actions independently, resulting in a performance that felt forced and less authentic.

## Discussion

In this section, we will illustrate some key findings from user study. One of the most common ways in which users interacted with the tool was by testing the improvisational skills of the AI by providing it with complicated inputs. For instance, users often tried to input contradictory emotions or unrelated keywords such as 'happy but sad', or 'chair, cloud, Napoleon' in the romance genre. Since the AI will always produce an output on any input regardless of the relationship of the inputs, users enjoyed exploring how an AI would make logical or amusing connections. On the other hand, a human might get writer's block when trying to produce a script using uncooperative keywords. From this finding, it is clear that this unlimited flexibility is one of the most valuable qualities of an AI scriptwriter.

Notable feedback indicated that Scribble.ai tends to write better stories about objects than it does about humans. This is perhaps due to the greater understanding we, as humans, have of our own species compared to AI. However, when it comes to objects, AI seems to have the upper hand due to its data-based knowledge, comprehending everything from the object's invention to its features. This allows the AI to incorporate elements into the story that humans might not observe or recognize. Some participants suggested that it would be interesting to utilize this phenomenon to craft stories centered around objects. They proposed humanizing these objects and weaving them into a narrative, thereby creating a more engaging theatrical piece. This insight provides an intriguing potential for AI-mediated storytelling with a shift towards non-anthropocentric narratives.

We received valuable feedback from professional actors and acting coaches who dedicated their time to testing our tool. As experts in improv activities, they have a strong sense of what to look for in an effective improv practice script. The most prominent critique was that the AI-generated scripts are too direct and specific and lack the usual ambiguity present in human-written scripts. A recurring comment received from amateur and professional actors alike is that they felt more comfortable working with AI-generated scripts that are more straightforward, as there is less needed to build background information and provide subtext. However, practice with those scripts is less effective since it limits their opportunities to train the actor's key role, to create the subtexts in the script. Despite the reduced scope for choices and decision-making, each actor was able to interpret the same script differently, showcasing individual creativity. Feedback from professionals suggests that the tool could be improved by generating scripts that are more open to interpretation, thereby allowing actors to utilize and develop their creative skills further.

One possible solution for the issues described above for the current version of the monologue function was proposed by Mark Rafael [35]. He suggested that an actor could generate a monologue with a given emotion, for example, happy, and swap the emotion that the character feels with its opposite, in this case, sad, while maintaining the same lines. This is more in line with the very human tendency to hide our emotions instead of showing them directly. This is also a good exercise since it can be challenging to build all the necessary subtext to make the scene presentable. For instance, acting out lines such as "I'm so happy to be here with you" while the character is feeling sad will force the actor to build a lot of background and context.

Since human-generated scripts often contain biases and recurring structures, actors can become overly familiar with these repetitive themes and conflicts. In contrast, AI scriptwriters draw from a vast range of authors, potentially reducing the inherent biases found in human writing. This capability allows AI to produce innovative ideas in a new storytelling format, offering a fresh style of playscript. We believe that users enjoyed and found value in collaborating with the tool due to its ability to deliver a unique acting experience. For instance, an acting coach noted that the tool could provide challenging prompts for students, offering a distinct style from the material he typically creates for them.

For future work, it would be beneficial to enhance the TLP model to generate scripts that extend beyond improv acting. Ultimately, the tool could also suggest screenplay directions, such as fade in/out, zoom in/out, and character motions. Our user testing indicated that the produced scripts lack the ambiguity typical of theatrical speech, often compensated for by body language and tone of voice. This led to scripts that are too direct and did not leave enough room for actors to get creative under the script. In the improvisational acting practice, it is particularly important that the script leaves room for diverse interpretations like a neutral scene. A neutral scene–often called a contentless scene– means a set of relatively ambiguous words or phrases written which can be interpreted in multiple ways [4]. Another avenue to explore could involve compiling a dataset of neutral scenes to train the model with. This would introduce ambiguity into the generated scripts and could even lead to a new feature that produces neutral scenes. To be considered successful, the TLP must be able to set and suggest directions for the scene using screenplay terminology as well as produce a script written with theatrical vernacular.

## Conclusion

This paper has introduced the new concept 'Theatrical Language Processing (TLP)', a new field at the intersection of linguistics, theatrology, computer science, and artificial intelligence dealing with the interactions between computers and theatrical language. We also developed and presented the first TLP tool, Scribble.ai.

Scribble.ai successfully generated creative and unique scripts for improv activity which are distinguished from scripts written by human writers. These scripts maintain enough similarity to human-written ones for easy comprehension while offering a distinctive use of keywords, often leading to more innovative outputs than those typically crafted by human writers. Scribble.ai also went through the user study which has confirmed the potential of AI-driven creativity support tools for actors. Additionally, Scribble.ai is beneficial as a productivity support tool, as actors using Scribble.ai can quickly generate custom practice scripts with ease. This will help overcome the challenge of accessibility and the limited amount of scripts.

The findings of this study underscore the significance of addressing the identified limitations and pursuing avenues for improvement in AI-driven creativity within theater and performance arts. By recognizing the need for more ambiguity in generated scripts and the potential for a specialized TLP model tailored to the nuances of film and theater, we lay the groundwork for enhancing the efficacy and adaptability of AI-powered tools in creative contexts.

Looking ahead, we envision continued collaboration and innovation at the intersection of AI and theater. Through ongoing refinement of TLP models, integration of user feedback, and exploration of emergent technologies, we strive to unlock new possibilities for storytelling, performance, and audience engagement. In doing so, we aim to foster a dynamic ecosystem where creativity flourishes and boundaries between human and machine creativity blur, ultimately enriching the cultural landscape and redefining the possibilities of theatrical expression.

## Acknowledgements

I sincerely thank Professors Eric Paulos, Björn Hartmann, and Yoon Bahk for their guidance and support from the early stages of this research. I am also grateful to Professors Ben Shneiderman and Terry Winograd, as well as ac-tor Mark Rafael, for their insightful interviews. Lastly, but mostly, I extend my deepest appreciation to Seon Kyung Jeong, theater director and playwright, for her invaluable assistance in conducting the experiment.

## Authors Biographies

Sora Kang is an interdisciplinary researcher at the intersection of Human-Computer Interaction and the arts. As a Ph.D. student at Seoul National University, she explores innovative applications of AI and LLM in creative practices. She holds a Master's degree in HCI from the University of California, Berkeley. Through her academic research and creative practice, she seeks to unlock AI's potential as a tool for artistic expression while employing artistic methods to develop smarter, socially intelligent AI systems.

Joonhwan Lee is a Professor in the Department of Communication at Seoul National University. He holds a Ph.D. in Human–Computer Interaction (2008) from Carnegie Mellon University. His research includes Human-Computer Interaction, human-ai interaction, ai journalism, social computing, situationally appropriate user interaction, and information visualization. He directs the Human-Computer Interaction+Design Lab (hcid.snu.ac.kr).